\documentclass{iopart}
\usepackage{graphicx}
\usepackage{dcolumn}
\usepackage{bm}
\usepackage{array}
\usepackage{multirow}

\newcolumntype {Z}{ >{\centering \arraybackslash }X}

\begin{document}

\title{The Effect of Hydrogen Adsorption on the Magnetic Properties of a Surface Nanocluster of Iron}

\author{Pavel F. Bessarab$^{1,2}$, Valery M. Uzdin$^{2,3}$, and  Hannes J\'onsson$^{1,4}$}
\address{$^1$Science Institute and Faculty of Physical Sciences, VR-III, University of 
Iceland, Reykjav\'{\i}k, Iceland}
\address{$^2$Department of Physics, St. Petersburg State University, St. Petersburg, 198504, Russia}
\address{$^3$National Research University of Information Technologies, Mechanics and Optics, St. Petersburg, 197101, Russia }
\address{$^4$Dept. of Applied Physics, Aalto University, Espoo,  FI-00076, Finland}

\begin{abstract}
The effect of hydrogen adsorption on the magnetic properties of an Fe$_3$ cluster immersed in a Cu(111) surface has been calculated
using densifty functional theory and the results used to parametrize an Alexander-Anderson model which 
takes into account the interaction of d-electrons with itinerant electrons.  A number of adatom configurations containing one to seven H-atoms were analyzed.  
The sequential addition of hydrogen atoms is found to monotonically reduce the total magnetic moment of the cluster with the effect being strongest when the
H-atoms sit at low coordinated sites. 
Decomposition of the charge density indicates a transfer of 0.4 electrons to each of the H-atoms from both the Fe-atoms and from the copper substrate, irrespective of adsorption site and coverage. 
The magnetic moment of only the nearest neighbor Fe-atoms is reduced and mainly due to
increased population of minority spin d-states. This can be modeled by increased indirect coupling of d-states via the conduction s-band in the Alexander-Anderson model.

\end{abstract}


\maketitle
\section{Introduction}
\label{sec:Introduction}

Hydrogen can strongly affect the magnetic properties of metals, in particular metal surfaces. This can be used to intentionally modulate magnetic properties by introducing hydrogen since hydrogen atoms can in many cases be adsorbed and desorbed relatively easily.  
Hydrogen can also be present inadvertently as an impurity and thereby affect measurements of magnetism.  Either way, it is of considerable importance to understand
the way in which hydrogen can affect magnetic properties of metals. 
The thermal stability of magnetic states of nanoscale islands on surfaces is strongly dependent on the size and shape of the islands 
and preferential adsorption of impurities on the island rim could affect this dependence.

In most cases, hydrogen loading has been found to lead to a monotonic change in the magnetic moment of large transition metal clusters and thin films as hydrogen concentration is increased. Surface magneto-optic Kerr effect experiments have been carried out in order to measure the effect of hydrogen on the magnetism of ultrathin films of Fe, Co and Ni on Cu(001)~\cite{mankey_93}. 
The magnetization of Ni and Co films has been found to be reduced upon
hydrogen adsorption, 
while for the Fe films, hydrogen enhances the magnetization slightly. Theoretical studies are consistent with these results~\cite{siegbahm_84,granucci_92,maca_03}. For example, {\it ab initio} calculations based on full-potential linearized augmented plane-wave method for a Ni multilayer on Cu(001) substrate have shown that hydrogen adsorption on the Ni film leads to a considerable decrease of magnetic moment in the upper Ni layers. However, magnetic properties of small, free standing iron, nickel and cobalt clusters, containing just a few atoms, respond to hydrogenation in a more complex way, in some cases showing oscillations in the magnetic moment upon successive addition of H atoms \cite{jones_04,ashman_03}. 

Long range effect of hydrogen adsorption on magnetic properties of atoms have also been reported.
While a full coverage of hydrogen on Co films was found to decrease the magnetization of surface atoms,
partial H-atom coverage led to a significant increase in the magnetic moment of surface Co-atoms not bonded to hydrogen, even leading to an overall enhancement of surface magnetism~\cite{gallego_10}. Another example of such long-range effect 
comes from studies of
Fe/V superlattices~\cite{remhof_07} using
element specific X-ray resonant magnetic scattering. A significant increase in the magnetic moment of the iron layers was found
when the vanadium layers were loaded with hydrogen even though the induced antiparallel moment of 
V-atoms at the Fe/V interface remained unaffected. This effect was explained using
model Hamiltonian calculations and is due to a redistribution of {\it d}-electrons between Fe and V atoms.
The introduction of hydrogen causes a shift of the {\it d}-band relative to the Fermi level and thus changes the exchange splitting~\cite{remhof_07}. 

We have chosen to study a system in between thin magnetic layers and small free standing clusters, 
namely a small iron cluster embedded in the surface of a non-magnetic metallic substrate. 
An embedded clusters rather than a cluster of adatoms on top of a substrate was chosen
because adatom clusters tend to be highly mobile via multi-atom concerted displacements (see for example ref.~\cite{Xu_PdClust}).  
Density functional theory (DFT) calculations are used to study the effect of hydrogen on the magnetic properties. Various configurations with the number of hydrogen atoms ranging from 1 to 7 at full saturation are considered.
The total magnetic moment of the iron cluster 
as well as the number of d-electrons, magnetic moment of Fe-atoms and the local density of states (LDOS)
is calculated as a function of the number of hydrogen atoms at the various types of
adsorption sites.
The effect of hydrogen adsorption is then reproduced with an Alexander-Anderson tight-binding model~\cite{AAmodel} with
parameter values chosen to reproduce the DFT results.
The AA model has previously 
been shown to describe adequately the magnetic states of clusters of 3$d$ transition metals on non-magnetic substrates.
For example, calculations of magnetic structure of supported transition metal clusters~\cite{uzdin_99,uzdin_01,uzdin_00} 
have been found to reproduce all the main features obtained within density functional theory~\cite{bergman_07,antal_08,lounis_07},
including the possibility of non-collinear ordering of magnetic moments and appearance of several stable magnetic states which are close in energy. The motivation for using such a model is to better elucidate the physical picture of the effect of hydrogen adsorption on the magnetic moment. 
The parametrized model can then also be used to estimate the effect of hydrogen adsorption on larger islands in systems that are too computationally demanding for the DFT calculations.

\section{The Simulated System}
\label{sec:SupIrTrim}

\begin{figure}
\begin{center}
\includegraphics[width=1.1\textwidth]{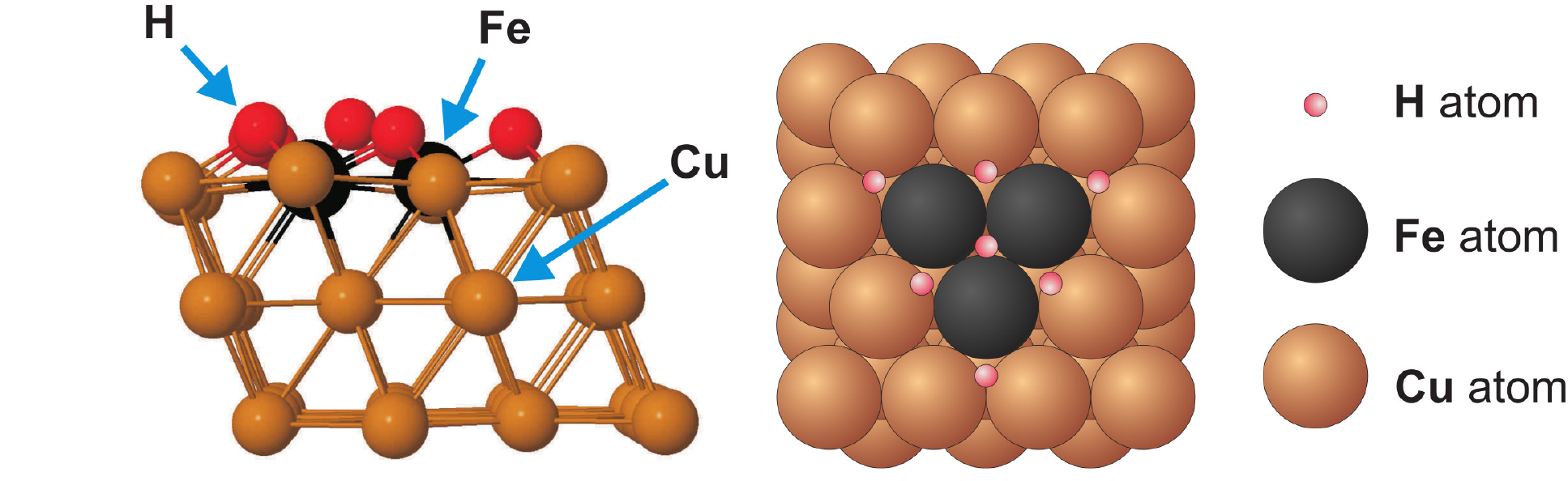}
\caption{Fe trimer cluster embedded into a Cu(111) surface. Up to seven H-adatoms can be adsorbed on the cluster.}
\label{fig:1} 
\end{center}
\end{figure}

The influence of hydrogen adsorption on magnetic properties of an iron cluster containing three atoms was calculated. 
A trimer was chosen because it is the minimal structure which contains the various types of sites for adatoms. 
The cluster is embedded into a Cu(111) surface and has the shape of an equilateral triangle. The Fe-Fe bond-lengths are fixed by the substrate and the geometry of the cluster is stable upon hydrogenation. 
The Cu substrate is represented by a
slab consisting of 33 atoms (see Fig.~\ref{fig:1}). There are three layers in the slab, each containing 12 metal atoms. The top two layers are allowed to move but the bottom one is kept fixed.
The DFT calculations made use of the Perdew, Burke and Ernzerhof (PBE)~\cite{PBE} approximation to the exchange and correlation functional.
A plane wave basis set with an energy cutoff of 273.25 eV was used to represent the valence electrons using the projector augmented wave (PAW) formalism~\cite{blochl}.
The Vienna {\it ab-initio} simulation package (VASP) was used in these calculations~\cite{VASP}.

Adsorption sites for the H-atoms were determined 
by placing the atoms at various sites on the cluster and carrying out energy minimization with respect to the atom coordinates.  Also, 
an initial structure with as many as 10 hydrogen atoms was set up and then heated to 750 K and annealed for 1.5 ps 
(1500 ionic moves) before finally minimizing the energy with respect to atom coordinates. 
During the annealing, three hydrogen atoms were desorbed from the system. 
The maximum coverage was thus determined to be 7 H-adatoms in an arrangement shown in Fig.~\ref{fig:1}. 

\begin{figure}
\begin{center}
\includegraphics[width=0.4\textwidth]{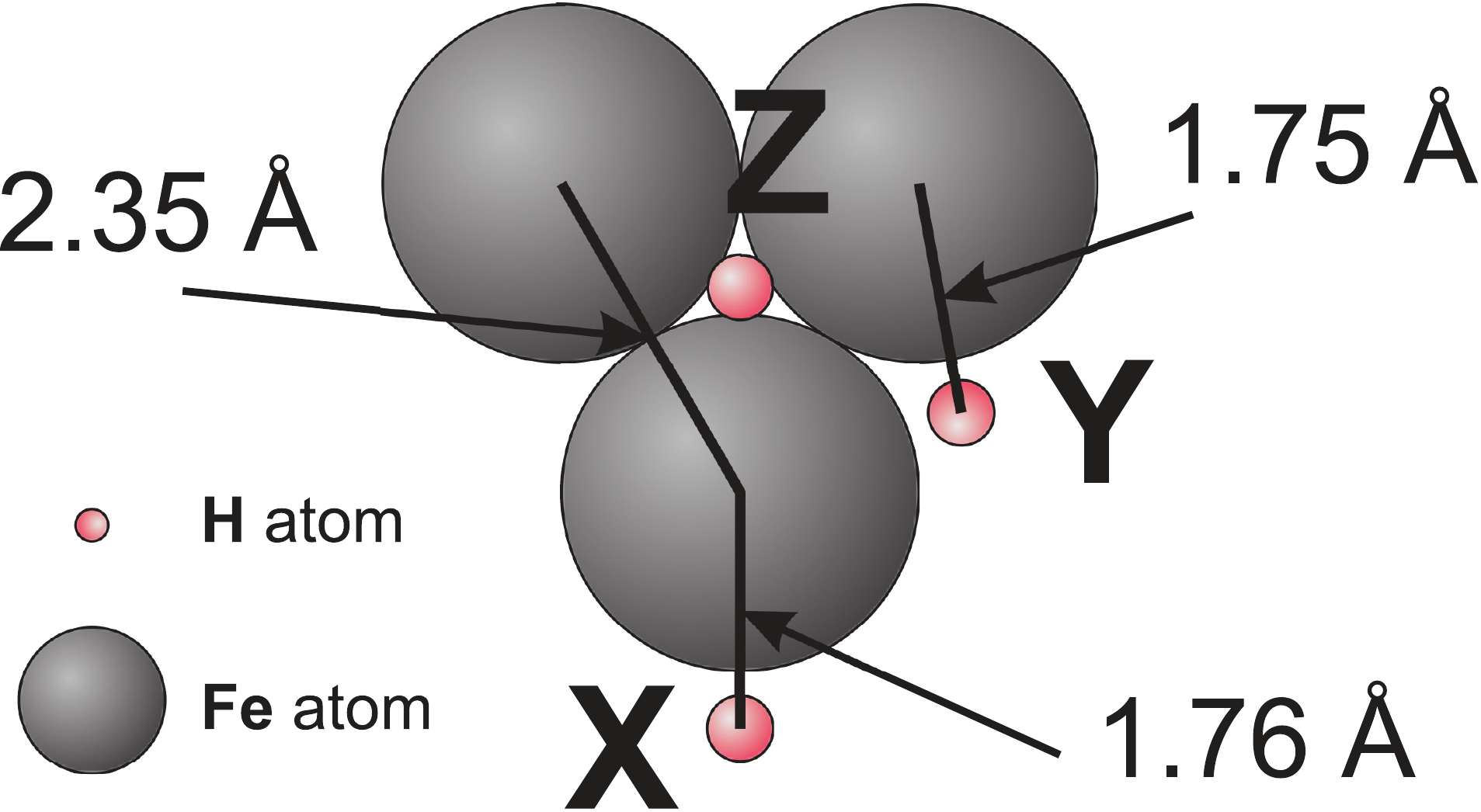}
\caption{On-top view of the Fe trimer cluster and the three possible sites for adsorbed H-atoms: X-, Y- and Z-site. The cluster is embedded in a (001) surface of a copper slab which is not shown.  The distance between Fe-H atom pairs is indicated.  }
\label{fig:2} 
\end{center}
\end{figure}

There are three types of adsorption sites differing in the number of nearest neighbor Fe-atoms. 
These are denoted as X-, Y- or Z-sites, corresponding to one, two and three nearest neighbor Fe-atoms, respectively (see Fig.~\ref{fig:2}). 
Up to three X-type, three Y-type and one Z-type adatoms may be present on the Fe trimer. Except for the Z-type adatom, distances between the H-atom and nearest Fe-atoms are nearly the same, 1.75 \r{A}. 
The adsorption of hydrogen was indeed found not to affect the geometry of the cluster. The Fe-Fe nearest neighbor distance turned out to be 2.35 \r{A} for all configurations of H-adatoms.

%

\section{The DFT Calculated Magnetization}

The total spin magnetic moment of the system was calculated using DFT for the various possible numbers and arrangement of the H-adatoms.
The calculated total magnetization of the various configurations is listed in Table~1
and the change due to H-atom adsorption shown in Fig. 3.

%

\begin{figure}
\begin{center}
\includegraphics[width=1.8\textwidth,angle=-90]{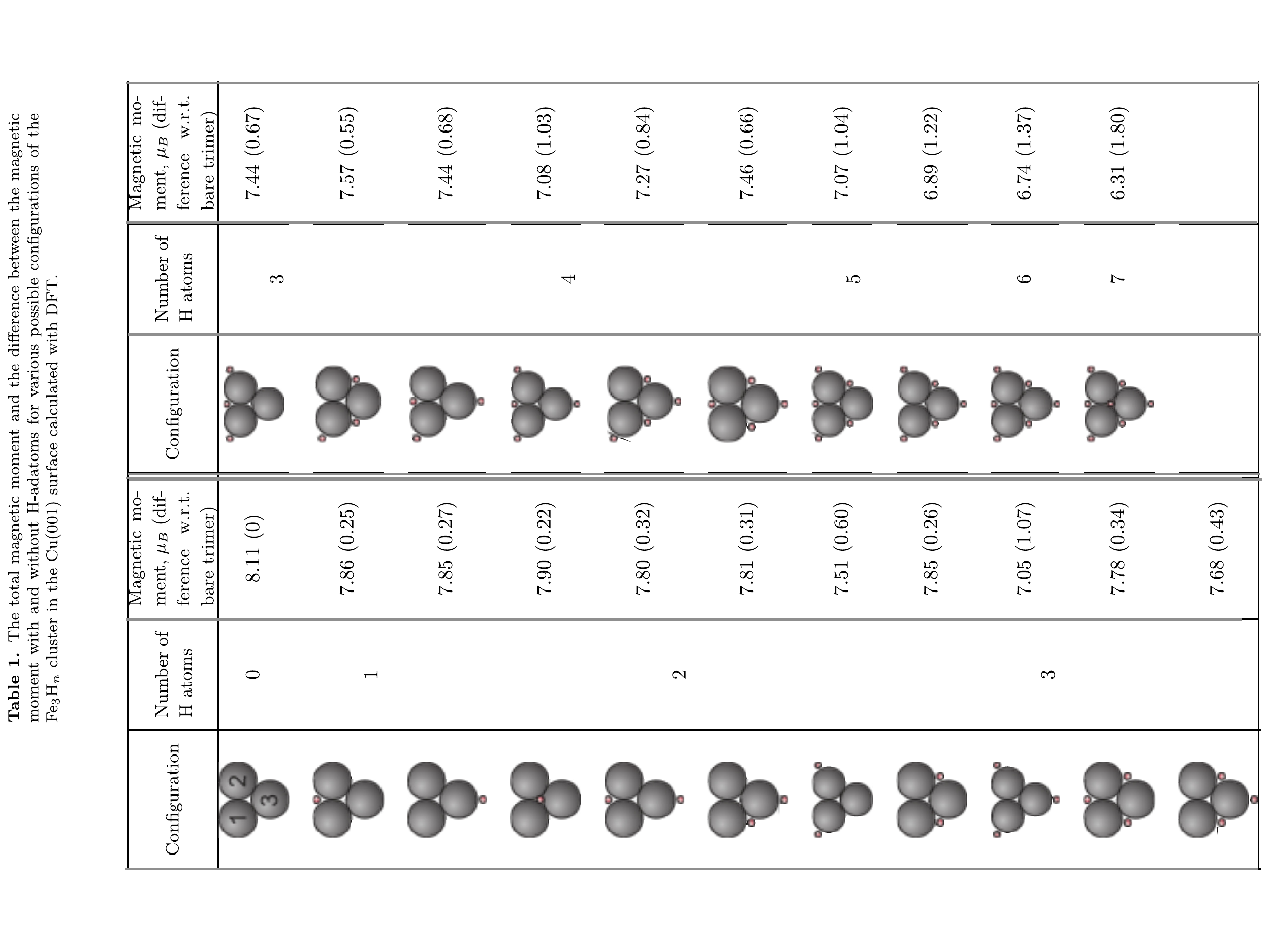}
\label{tab:1} 
\end{center}
\end{figure}
%


Two trends are evident: 
(1) The total magnetic moment decreases as H-adatoms are introduced. 
(2) For a given number of H-adatoms 
the strength of the effect depends on the number of Fe-neighbors of the H-adatom, 
an X-type adatom lowering the magnetic moment more than a Y-adatom, which in turn lowers it more than a Z-type adatom.
The effect of X-type adatoms is approximately twice as great as that of Y-type adatoms. 
The calculated change in the magnetic moment is shown in Fig.  3.

This decrease in the value of the total magnetic moment upon hydrogen adsorption is analogous to what has been observed in thin
magnetic layers and large magnetic clusters~\cite{mankey_93,siegbahm_84,granucci_92,maca_03}
as opposed to the oscillatory change observed for 
some small free standing transition metal clusters, such as Fe$_n$ and Co$_n$~\cite{jones_04,ashman_03}. 
The electronic structure of bulk materials and surfaces can be significantly different from that of small clusters.
Charge decomposition and changes in the LDOS 
can help gain an understanding of these effects. 
This type of analysis is presented in the next section.

\begin{figure}
\begin{center}
\includegraphics[width=0.7\textwidth]{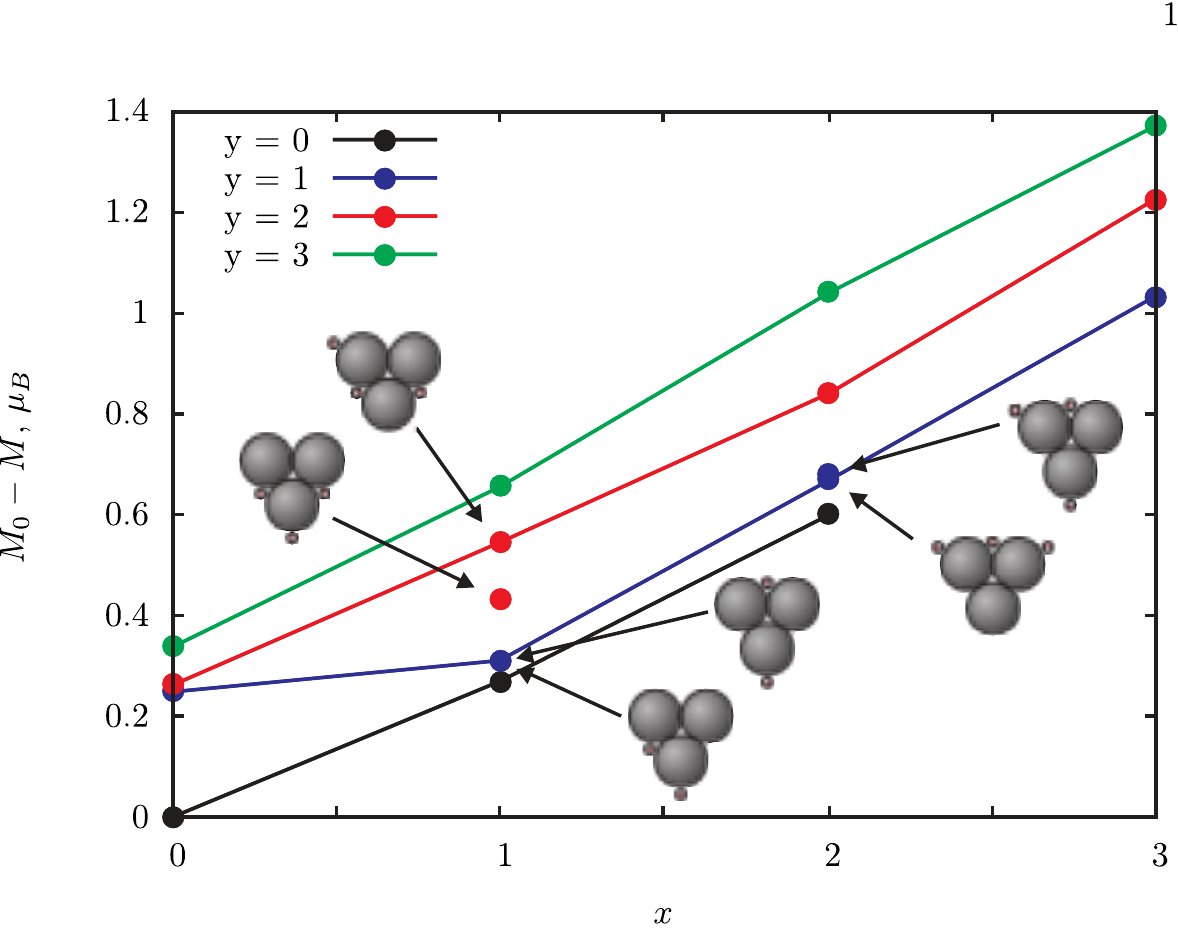}
\caption{DFT calculated change in the total magnetic moment of the Fe$_3$H$_n$ cluster as a function of the number of H-adatoms of type X for a given number of Y-type H-adatoms (see legend and insets). 
}
\label{fig:3} 
\end{center}
\end{figure}
%


\section{Analysis of charge density and DOS}
\label{sec:LocEffAdsHAto}
In order to gain insight into the changes taking place in the electronic structure upon adsorption of hydrogen, we have
carried out a charge decomposition analysis using the Bader definition of atomic regions~\cite{Bader}. A fast, grid based algorithm including LDOS analysis was used~\cite{Henkelman06,Gudmundsdottir12}.
The results, listed in Table 2, show that
each hydrogen atom attracts the same number of electrons, about 0.4, irrespective of the site. No spin density is found
within Bader regions associated with the H-atoms.  The number of electrons associated with Fe-atoms only changes for the nearest neighbors of the H-adatoms. 
For example, an H-adatom at a Y-site formed by two Fe-atoms neither changes the total number of electrons nor the number of unpaired electrons associated with the third Fe-atom. The same holds for an H-adatom at a X-site, it changes significantly the electronic structure of the nearest Fe-atom.
The total number of electrons at the Fe-atom decreases by $\sim$0.1 and magnetic moment decreases by $\sim$0.25$\mu_B$, but the number of electrons at the other two Fe-atoms in the cluster is not affected significantly. An H-adatom at a X-site reduces the number of unpaired electrons more strongly than an adatom at a Y-site. 


\begin{figure}
\begin{center}
\includegraphics[width=1.2\textwidth]{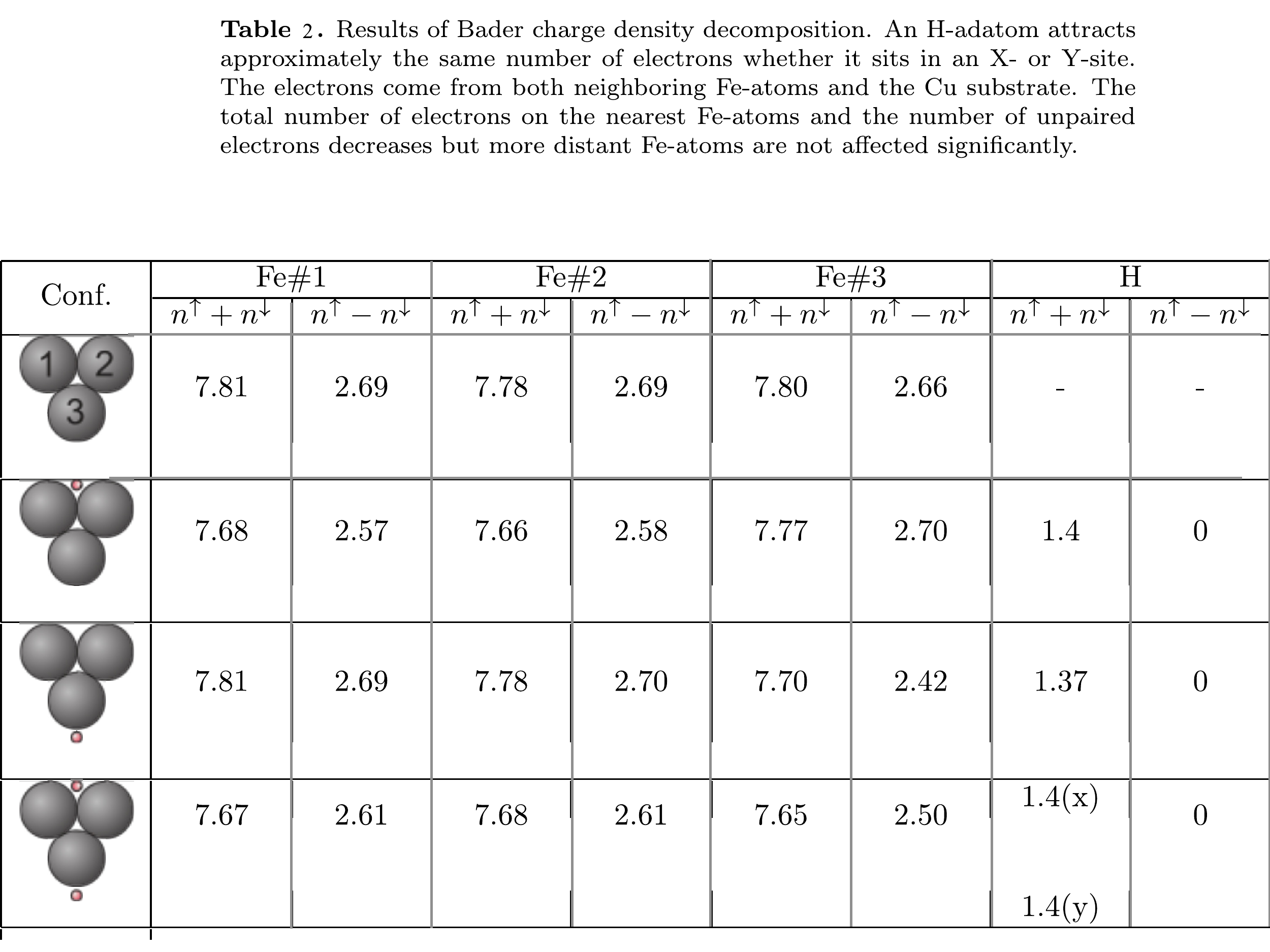}
\label{table:2} 
\end{center}
\end{figure}
%

%
\begin{figure}
\begin{center}
\includegraphics[width=0.9\textwidth]{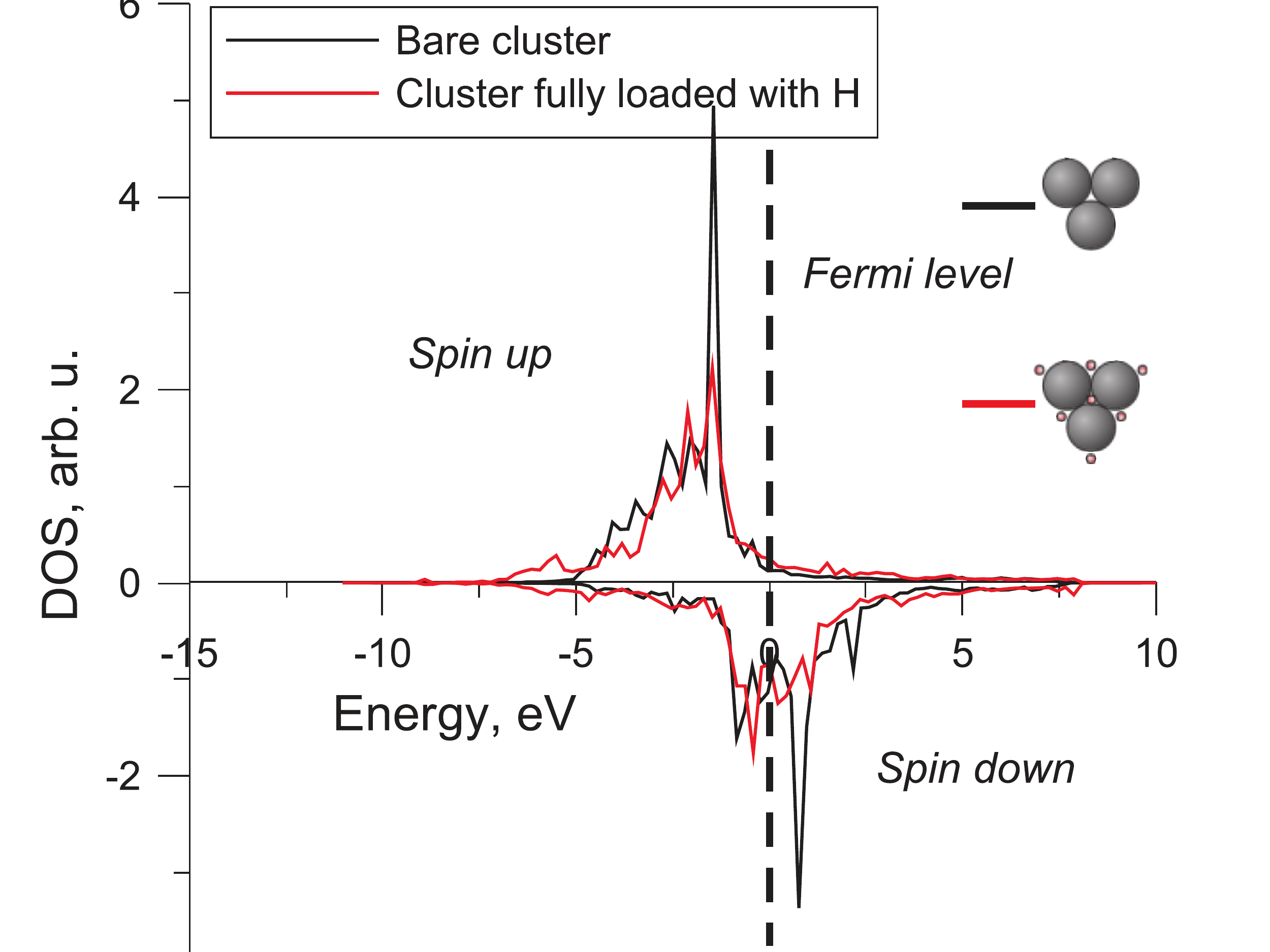}
\caption{DFT calculated local density of states for {\it d}-electrons in a Bader region of an Fe-atom in a Fe$_3$ cluster embedded in a Cu(001) surface, 
with and without H-adatoms (see inset).
The black line corresponds to the bare Fe cluster. The red line corresponds to a cluster with full coverage of hydrogen, see insets. 
The Fermi level is shown with a dashed line.
The minority-spin (spin-down) states 
get filled to a larger extent in
the presence of hydrogen, thus reducing the 
difference in the number of spin-up and spin-down electrons, and thereby the magnetization.
 }
\label{fig:4} 
\end{center}
\end{figure}

The DFT calculated LDOS for {\it d}-electrons in the Bader region of an Fe-atom in a cluster with no and full hydrogen coverage is shown in Fig. 4.  
The interaction of the {\it s}-state of the H-atom with the {\it d}-band of the Fe-atoms changes the band structure leading to the considerable changes of both majority- and minority-spins. 
The majority-spin states are almost all below the Fermi energy, and their redistribution with hydrogenation does not lead to significant changes in their occupation. Integration of the majority-spin LDOS up to the Fermi level decreases by only 0.1 in the presence of hydrogen. 
The integral of minority-spin states up to the Fermi energy is, however, changed much more, it increases from 1.6 to 2.1 upon hydrogenation. This results in reduced magnetic moment of the Fe-atoms. 
H-adatoms also induce changes in deeper states but these are not relevant for the magnetic properties.
The insight gained from this analysis is used in the next section to parametrize a model which can reproduce the observed changes in
the magnetization of the cluster.

\section{Analysis with a model Hamiltonian}
\label{sec:AnModHam}

In order to analyze the effects of hydrogen adsorption and to enable calculations of much larger systems, the DFT results are used to parametrize an Alexander-Anderson
model for the system~\cite{AAmodel}.
The model describes two electronic bands: 3{\it d}-electrons localized on the Fe-atoms of the cluster 
and itinerant {\it s}({\it p}) electrons.
The Hamiltonian of the system is written as
\begin{equation}
\label{eq:1}
\begin{array}{l}
H=\sum_{\bf{k},\sigma}E_{\bf k}\hat{n}_{\bf k\sigma}+ \sum_{i,\sigma}E_{0i}\hat{n}_{i\sigma}+ \sum_{{\bf k},i,\sigma}V_{i\bf k}\hat{d}_{i\sigma}^{\dag}\hat{c}_{\bf k\sigma}\vspace{10pt}\\ \hspace{2em}+\sum_{i\ne j,\sigma}V_{ij}\hat{d}_{i\sigma}^{\dag}\hat{d}_{j\sigma}+ \frac{1}{2}\sum_{i,\sigma}U_i\hat{n}_{i\sigma}\hat{n}_{i-\sigma}+ h.c.\\
\end{array}
\end{equation}
where $h.c.$ stands for Hermitian conjugate.
Here $\hat{d}_{i\sigma}^{\dag}(\hat{d}_{i\sigma})$ and $\hat{c}_{\bf k\sigma}^{\dag}(\hat{c}_{\bf k\sigma})$ are creation (annihilation) operators for {\it d}-electrons with spin $\sigma=\pm$ localized on site {\it i} and itinerant {\it s}({\it p})-electron with quasimomentum {\bf k} respectively; 
$\hat{n}_{i\sigma}=\hat{d}_{i\sigma}^{\dag}\hat{d}_{i\sigma}$, $\hat{n}_{\bf k\sigma}=\hat{c}_{\bf k\sigma}^{\dag}\hat{c}_{\bf k\sigma}$ are corresponding occupation number operators. 
The energy of non-interacting {\it s}({\it p}) electrons, $E_{\bf k}$, and {\it d}-electrons, $E_{0i}$, hybridization parameters, $V_{i{\bf k}}$, hopping parameters, $V_{ij}$, and Coulomb repulsion on site, $U_i$, are spin independent. The last term in the Hamiltonian in eqn.(\ref{eq:1}) is included within a mean field approximation
\begin{equation}
\label{eq:2}
U_i\hat{n}_{i\sigma}\hat{n}_{i-\sigma}
\approx
U_i\hat{n}_{i\sigma}\langle\hat{n}_{i-\sigma}\rangle + U_i\langle\hat{n}_{i\sigma}\rangle\hat{n}_{i-\sigma} - U_i\langle\hat{n}_{i\sigma}\rangle\langle\hat{n}_{i-\sigma}\rangle,
\end{equation}
where $\langle\hat{n}_{i\sigma}\rangle$ denotes the average value of an occupation number. These numbers are found self-consistently within a Green function method.

Only the {\it d}-electrons are considered explicitly in the calculations. Therefore, only the three last terms of the Hamiltonian are included while the influence of the {\it s}({\it p})-electrons is taken into account via the renormalization of the phenomenological model parameters. In particular, {\it s}-{\it d}-hybridization leads to the appearance of a non-zero width, $\Gamma$, of the {\it d}-level which is independent of the hopping parameters, $V_{ij}$. The model has adjustable parameters which in the present case are deduced from the results of DFT calculations.

The Fe atoms are characterized by two dimensionless phenomenological parameters. The first determines the 
energy level of non-interacting {\it d}-electrons relative to the Fermi energy in units of the width of the {\it d}-level, $\Gamma$,
which arises due to {\it s}-{\it d} hybridization,
${E_i^0}/{\Gamma}$. The second parameter represents the Coulomb repulsion on a site, again 
relative to 
$\Gamma$,  ${U_i}/{\Gamma}$. These parameters are determined by reproducing the DFT calculated magnetic moment and number of {\it d}-electrons 
of a single Fe-atom in the Cu(001) substrate. The values obtained are ${U}/{\Gamma}=13$ and ${E^0}/{\Gamma}=-12$.

The model also includes a dimensionless hopping parameter ${V_{ij}}/{\Gamma}$, 
describing
direct coupling of {\it d}-electrons of atoms {\it i} and {\it j} as well as the indirect {\it d}-{\it s}({\it p})-{\it d} coupling via the conductivity band.
Since both contributions decrease with the distance between the atoms, hopping parameters are a measure of the interatomic distances and relate to the geometry of the cluster. 
Here, only one value of this parameter needs to be chosen for the hydrogen free island.  
We have used the value ${V_{ij}}/{\Gamma}=0.39$ in order to reproduce the magnetic properties obtained in the DFT calculation.

The model for the embedded Fe$_3$ cluster, where all atoms are equivalent, therefore, contains only three dimensionless parameters:
${E^0}/{\Gamma}$, ${U}/{\Gamma}$ and ${V}/{\Gamma}$. Any one of these three parameters could in principle change upon hydrogenation. 
We will, however, search for the simplest change in model parameters that can reproduce reasonably well the DFT results discussed above.


\subsection{The effect of hydrogen adsorption in AA model}
\label{sec:IncrInHopPar}

The LDOS shown in Fig. 4 shows that there is not an appreciable change in the width of the {\it d}-band
upon hydrogenation.  The parameter $\Gamma$ is, therefore, taken to be unaffected. 
The position of the {\it d}-level relative to the Fermi energy, ${E_i^0}/{\Gamma}$, and the Coulomb repulsion of {\it d}-electrons localized on 
site {\it i}, ${U_i}/{\Gamma}$, appear from the DFT calculations also not to be affected much by the presence of H-adatoms.  
The presence of the hydrogen {\it s}-state will, however, change the hopping parameters ${V_{ij}}/{\Gamma}$. Indeed, 
hydrogen attracts some electrons thus leading to a local increase in electron density at the adsorption site. This in turn increases indirect coupling between {\it d}-electrons of neighboring Fe atoms through the conductivity band. 

Given the localized effect of the H-adatoms, we have chosen to let an H-atom at a Y-site increase the hopping parameter between the
two Fe-atoms it is bonded to.  An H-atom at an X-site increases the two hopping parameter between the neighboring Fe-atom and each of its two neighbors.
An increase in the hopping parameters ${V_{ij}}/{\Gamma}$ is indeed found to decrease the magnetic moment of the Fe-atoms in the cluster. 
A rough fit to the DFT data gives an increase, $\Delta V$, in the hopping parameter, $V_{ij}$, for each H-atom bound to either one or both Fe-atoms $i$ and $j$, with $\Delta V/V =  0.25$.
The calculated magnetization with these parameters in the AA model is compared with the DFT results in Fig. 5.

\begin{figure}
\begin{center}
\includegraphics[width=0.7\textwidth]{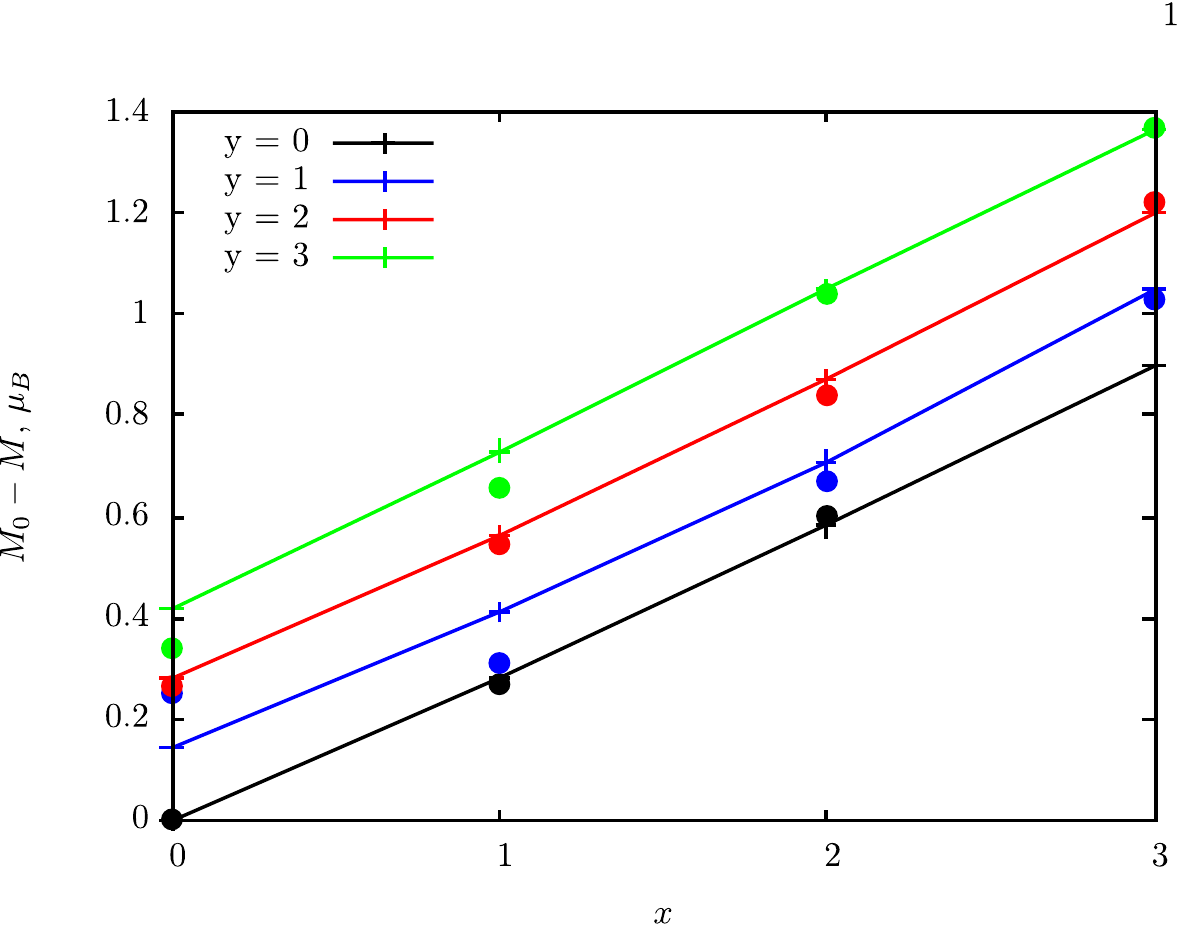}
\caption{The total magnetic moment of Fe$_3$H$_n$ cluster as a function of the number of H-adatoms in X- and Y-sites. 
Crosses correspond to calculations with the AA-model~\cite{AAmodel} 
where the adsorption of each H-adatom increases the value of hopping parameters of the neighboring Fe-atoms by $\Delta V$ where $\Delta V/V =  0.25$. 
Circles show results of DFT calculations.}
\label{fig:5} 
\end{center}
\end{figure}

The correspondence between the results of the AA-model and the DFT calculations is good especially considering the simplicity
of the parameter adjustment.
However, significant deviations occur when only a few H-adatoms are present. 
Most likely,  other parameters of the model should also be adjusted slightly to reproduce more closely the effect of the hydrogen adsorption. 
A more elaborate fitting procedure could be carried out, but here we focus on the larger trends and relatively high coverage.

%
%
\section{Conclusion}
\label{sec:conclusion}

The DFT calculations presented here have shown that the adsorption of H-atoms on a Fe-atom cluster embedded in a Cu(111) substrate reduces the total magnetic moment of the cluster. 
The strength of the effect depends on the adsorption site of the H-atom, in that an H-atom bound to a single Fe-atom leads to a stronger reduction in magnetic moment than an H-atom bound to two Fe atoms. 
The Bader decomposition of the charge density 
shows that an H-atom draws ca. 0.4 electrons from the Fe-atom(s) they are bound to as well as the 
substrate.
An analysis of the LDOS of {\it d}-electrons for individual Fe-atoms shows that the reduction of 
magnetic moment is primarily due to an increase in the occupation of minority-spin states, which are near the Fermi level.

The DFT results can be reproduced quite well with a simple, AA model where the adsorption of H-atom
increases the indirect coupling between {\it d}-states via the conduction band.
A 25\% increase in the relevant $V_{ij}$ hopping parameters of the model upon hydrogen adsorption 
gives good agreement with the magnetic moments of the Fe-atoms obtained from the DFT calculations. 

Once the AA model has been parametrized, it can be used to simulate much larger systems than can be studied by DFT calculations, even systems with thousands of atoms. Furthermore, a recently developed harmonic transition state rate theory for 
magnetic transitions~\cite{bessarab_12} can be used to analyze size and shape dependence of thermal stability of magnetic states of 
nano-islands on surfaces, as has been demonstrated for Fe islands on W(110)~\cite{bessarab_13}. These kinds of calculations can be
performed using the AA model~\cite{bessarab_13b} and the effect of hydrogen adsorption on thermal stability of 
magnetic states assessed for a wide range 
in island size and shape.  This will be the subject of future work.


\ack

We thank Dr.~Andri Arnaldsson for help with the DFT calculations.  This work was supported by the Icelandic Research Fund, the Nordic Energy Research fund and the Academy of Finland.


\end{document}